\documentstyle[12pt,epsfig]{article}
\newcommand{\m}{\medbreak}

\newcommand{\no}{\noindent}
\newcommand{\EQ}{\begin{equation}}
\newcommand{\eq}{\end{equation}}
\newcommand{\EQA}{\begin{eqnarray}}
\newcommand{\eqa}{\end{eqnarray}}

\newcommand{\AR}{\renewcommand {\arraystretch}{1.5}
\begin{array}{l}}
\newcommand{\bAR}{\renewcommand {\arraystretch}{2}
\begin{array}{l}}
\newcommand{\ARc}{\renewcommand {\arraystretch}{1.5}
\begin{array}{c}}
\newcommand{\bARc}{\renewcommand {\arraystretch}{2}
\begin{array}{c}}
\newcommand{\ar}{\end{array} \renewcommand {\arraystretch}{1}}

\newcommand{\ET}{\mbox{$E_T\ $}}

\newcommand{\ALLPV}{\mbox{$A_{LL}^{PV}\ $}}

\newcommand{\r}{\rightarrow}

\newcommand{\Z}{$Z^{\circ}\ $}
\newcommand{\ZP}{$Z'\ $}

\begin{document}
\begin{titlepage}
\vspace{0.2in}
\vspace*{1.5cm}
\begin{center}
{\large \bf A light leptophobic Z' in polarized hadronic collisions
\\} 
\vspace*{0.8cm}
{\bf P. Taxil} and {\bf J.-M. Virey}  \\ \vspace*{1cm}
Centre de Physique Th\'eorique$^{\ast}$, C.N.R.S. - Luminy,
Case 907\\
F-13288 Marseille Cedex 9, France\\ \vspace*{0.2cm}
and \\ \vspace*{0.2cm}
Universit\'e de Provence, Marseille, France\\
\vspace*{1.8cm}
{\bf Abstract} \\
\end{center}
Theoretical and phenomenological arguments are in favor of an elusive new neutral
vector boson \ZP with a relatively low mass, 
chiral couplings to ordinary quarks and
whose couplings to ordinary leptons are suppressed (leptophobia). 
We point out that this new particle could induce some
parity violating spin asymmetries which could be measured soon
at the Brookhaven Relativistic Heavy Ion Collider (RHIC), running part of 
the time as a polarized hadronic collider. \\

\vfill
\begin{flushleft}
\m\no
Number of figures : 3\\

\m\no
July 1998\\
CPT-98/P.3667\\
\m\no
anonymous ftp or gopher : cpt.univ-mrs.fr

------------------------------------\\
$^{\ast}$Unit\'e Propre de Recherche 7061
 \\
E-mail : Taxil@cpt.univ-mrs.fr ; Virey@cpt.univ-mrs.fr
\end{flushleft}
\end{titlepage}

\section{Introduction}
\indent
\m
One of the simplest extensions of the Standard Model (SM) is the addition of an extra
$U(1)'$ gauge factor to the $SU(3)\times SU(2) \times U(1)$ structure.
If the symmetry breaking occurs at a scale not far from the electroweak scale, this
leads to the existence of a new neutral gauge boson \ZP at a mass
accessible to forthcoming experiments.\\
In recent years, some
discrepancies with respect with the Standard Model expectations observed at LEP 
and/or
FNAL have triggered a lot of studies involving a possible
\ZP whose couplings to ordinary leptons could be very small (leptophobia).\\
Although the discrepancy in the $Z \r b\bar b$ sector has not completely
disappeared, the improved agreement between the latest data \cite{LEPEWWG}
and the SM expectations
has weakened the phenomenological motivations for that object (see e.g.
\cite{Mangano}).
However, the interest on such a leptophobic \ZP goes far beyond the
tentative for explaining the LEP and CDF data. 
Various models, based on Grand Unification and/or Superstring theory, display
a parameter space which allows or favors a new neutral gauge boson
with very small couplings to leptons. On the other hand, leptophobia still 
represents an attractive possibility allowing the existence of new physics
at an accessible energy scale without any contradiction with the present data.
\m 
It has been emphasized in refs. 
\cite{CveticLang1,CveticLang2,Lykken2,LangackerWang} that some 
non minimal SUSY models with an additional $U(1)'$ imply the presence of
a relatively light \ZP : $M_{Z'} < 1\,$ TeV/$c^2$. Indeed, a class of models, driven
by a large trilinear soft SUSY-breaking term, prefers the range 
$M_Z \leq M_{Z'} \leq 400 \,$GeV/$c^2$ along with a very small mixing with the
standard \Z. This opens some very interesting possibilities for
phenomenology provided that this new vector boson displays leptophobic couplings to
remain compatible with present data.
\m
A leptophobic \ZP is particularly elusive as far as conventional direct searches 
via Drell-Yan pair production at a $p\bar p$ collider are concerned.
It is of some importance to consider other possible manifestations of this new
boson. 
\m
If produced in hadronic collisions, the new \ZP can decay into exotic
fermions whose presence is necessary for anomaly cancellations \cite{Rosner}.
Such decays can yield some anomalous events which are, however, difficult to
interpret.\\
An other interesting feature has been pointed out by Rosner \cite{Rosner} :  the
direct \ZP couplings to quarks (which can be generation dependent or not) 
often break chiral
symmetry at a variance with QCD which is left-right symmetric. This can yield a
substantial forward-backward asymmetry  $A_{FB}$ in $p\bar p \r Z' \r f\bar f$ 
events. Such measurements are particularly difficult in the absence of outgoing
leptons (leptophobia) since it is mandatory to measure
the charge of the outgoing particle or jet. 
Moreover, in some case like in the model
described in \cite{Rosner}, $A_{FB}$ = 0 in the production of down(bottom)-type
quarks which forbids the opportunity of using $b$-tagging.
\\
There is an other way to be sensitive to chiral couplings, namely
the measurement of a parity-violating (PV) spin effect in polarized 
hard collisions of hadrons, in particular in the production of 
jets \cite{TannenbaumPenn}. In fact, within
two years, the RHIC Spin Collaboration (RSC) \cite{RSC} will start running the
Relativistic Heavy Ion Collider at  Brookhaven National Laboratory part of the time
in the  $\vec p \vec p$ mode, with a high (70\%) degree of polarization and with a
very high luminosity ${\cal L}\ =\, 2. 10^{32} cm^{-2}s^{-1}$. The physics goals of
the collaboration have been discussed in a recent workshop \cite{RIKENBNL}.
\m
As soon as one can find some arguments for 
{\it -i)} an extra $U(1)'$ gauge factor
leading to a relatively light \ZP , {\it -ii)} a parameter space which favors
some very weak couplings of this \ZP to conventional leptons,
{\it -iii)} a model in which an asymmetry in the left and right-handed couplings to
light quarks is preferred or at least allowed, then it is extremely interesting to
explore the consequences on the spin observables which will be measured soon, with
a high degree of precision, by the RHIC Spin Collaboration.\\

In the following section we will summarize the theoretical
motivations for a light and leptophobic \ZP with chiral couplings to quarks. 
We also give the parameter space which 
follows from the models we consider, along with the
present experimental direct limits on the mass of the \ZP 
and the strength of its coupling to light quarks.
The effects of the new amplitudes on the spin
asymmetries in inclusive jet production at the polarized RHIC collider are presented
in Section 3, along with the bounds on the parameter space which could be obtained.
 We give our conclusions in Section 4.

\section{Theoretical motivations and the parameter space}

\m
We will consider a new neutral vector boson \ZP which couples to $u$- and 
$d$- type quarks with the following structure  : 
\EQ
\label{lag}
{\cal L}_{Z'} = \kappa {g\over 2 \cos \theta_W} Z'^{\mu}{\bar q} \gamma_\mu[ C^q_{L}
(1 - \gamma_5) \; +\; C^q_{R} (1 + \gamma_5) ] q
\eq \no
for each given quark flavor $q$, the parameter $\kappa = g_{Z'}/g_Z$ being
of order one. 
\m
We will not discuss any more the phenomenological models with strong
couplings and $M_{Z'} = 1$TeV/$c^2$  which were built \cite{AltaChiap} 
to explain the old LEP (+CDF) data. 
Spin asymmetries at RHIC for these particular
models have been already discussed in Ref. \cite{TVZprime}.
\\ 
A theoretically motivated leptophobic \ZP with chiral couplings is
present in  many string inspired  models : in particular we will consider a version
due to Lopez and Nanopoulos \cite{LopezNanopoulos} (model A)
of the flipped-SU(5) model \cite{AEHN} and a generalized $E_6$
model with kinetic-mixing \cite{BKMR12} (model B). 
\m
- In model (A), the particle content of the standard model  is contained in
the representations : $F = ({\bf 10},{1 \over 2}) = \{ Q,d^c,\nu^c \} $
, $\bar f = ( \bar {\bf 5}, -{3\over 2}) = \{ L,u^c \} $ , 
$l^c = ( \bar {\bf 1}, {5 \over 2}) = \{ e^c \} $.
\m
Leptophobia is achieved  when $\bar f$ and $ l^c$, which contain
the standard model leptons, are uncharged under the new $U(1)'$.
As a consequence (see \cite{LopezNanopoulos}) the quark couplings
verify $C_L^{u} = C_L^d = - C_R^d = 1/(2\sqrt 3)$ and $C_R^{u} = 0$.
Therefore parity is maximally violated in the up-quark sector and is conserved
in the down-quark sector since the resulting coupling is purely axial.
\m
- Model (B) goes beyond the traditional parametrization in terms of the $U(1)$'s
combination in $E_6$ models (for a review of $E_6$ models
see \cite{Hewett1}). In fact, it was noticed some years ago \cite{delA2} that it 
was possible within $E_6$ models,  to find a particular combination
of charges leading to a suppression of the \ZP couplings to ordinary leptons
(see \cite{Rosner} for a recent discussion).
In Ref.\cite{BKMR12} one considers that 
kinetic mixing  (KM) between the field strength of weak hypercharge $B_{\mu\nu}$
and the field strength of the new $U(1)'$ occurs. In the interesting case of
KM between  the so-called $U(1)_{\eta}$ and $U(1)_Y$, one gets 
leptophobia when,  for each fermion $f$,  the \ZP couples to a combination 
$Q'^f = Q^f_{\eta} + \delta Y_W$ with $\delta = -1/3$.
This yields the quark couplings : 
$C_L^{u} = C_L^d =  C_R^d = -{1\over 2} C_R^{u} = - {5\over 18} \sin \theta_W$. 
\\
Note that to maintain $\kappa$  of order one in eq.(\ref{lag}), the usual GUT factor
${\sqrt {5\over 3}} \sin \theta_W$ has been included in the $C^q_{L,R}$'s couplings.
\m
Since a leptophobic \ZP appears in several other string derived models
\cite{Lykken2,FaraggiMasip} it is valuable to consider a more general
situation where the \ZP couplings to ordinary quarks are less constrained.
We will nevertheless assume $SU(2)_L$ invariance which imposes
$C_L^{u} = C_L^d \equiv C_L$. Then, we are left with three parameters :
$C_L$, $C_R^{u}$ and $C_R^d$. Of course in some models the couplings are generation
dependent : since we will be interested in ordinary jets, this point is
not relevant for our analysis.\\
Non-SUSY
leptophobic models have also been constructed \cite{Agashe,GG96}. 
Again there is a large freedom in the precise values for the \ZP couplings
to light quarks. 
We will consider for illustration
the special case introduced in Ref. \cite{GG96} where the
\ZP is purely right-handed (model C) : $C_L = 0$, $C_R^{u} = C_R^d = 1/3$ (to
be consistent with our notations we have divided the value given in \cite{GG96}
by a factor 3 to keep the corresponding $\kappa \approx 1$).
\m
The constraints on models (A) and (B) which are
coming from updated electroweak data have been studied very recently
\cite{Hagiwara12}. It turns out that for a small value of the \Z - \ZP mixing
angle $\xi$ the \ZP mass remains essentially an unconstrained parameter
in case of pure leptophobia. From the theoretical side a very small
mixing angle is natural in the scenario advocated in \cite{CveticLang2}.
One gets the same behaviour when the new $U(1)'$ does not mix with the SM gauge group
like within a class of models which can include model(A) \cite{LopezNanopoulos}.
As a consequence we have neglected the \Z - \ZP mixing angle $\xi$ which is in
principle present in all the $C_{L,R}^q$ couplings and whose effect
on the calculations described below would be very tiny.\\
 Direct constraints on $M_{Z'}$ are coming from $p-\bar p$ collider experiments
analyzing the dijet cross section  : UA2 \cite{UA2}, CDF \cite{CDFWjets2} and
D0 \cite{D0jets}. These limits are usually displayed in term of the 
so-called "sequential standard model \ZP" with $\kappa=1$. It is not difficult
to perform an extrapolation  for a reasonable range of $\kappa$ values and for
a larger class of \ZP models with various $C_{L,R}^q$ couplings. We have displayed
the limits in Fig.1 for models (A) and (B) assuming a 100\%
branching ratio of the \ZP into ordinary jets. One can notice that
the \ZP mass is not constrained in these models as soon as $\kappa < 0.95$. Also, 
and this is true in any leptophobic model, 
some windows are present around $M_{Z'}$ = 300 GeV$/c^2$ and below 
$M_{Z'}$ = 100 GeV$/c^2$.\\
Therefore, one can see that present data
are not excluding a \ZP as light as the one which is advocated in recent
papers based on weak-scale supersymmetry
\cite{CveticLang1,CveticLang2,Lykken2,LangackerWang}.
As discussed by Lykken \cite{Lykken2} this scenario could take place for model (A).
This is also the case for model (B), according to Refs.\cite{LangackerWang}, if
additional matter from an extra {\bf 78} representation of $E_6$ is involved.
Finally, we will also consider the very special case where
the \ZP is degenerate in mass with the SM \Z. This can be accepted in the framework
of the trilinear scenario of Ref. \cite{CveticLang2} and it was stressed in 
Ref.\cite{CaravagliosRoss} that such a leptophobic boson could explain the 
apparent discrepancy
between the LEP and SLAC values of $\sin^2\theta_W$.
To summarize, we will focus on a mass range $M_Z \leq M_{Z'} \leq 400$ GeV which
could give spectacular effects in PV spin asymmetries at RHIC. \\

\section{Calculation and results}
At RHIC, running in the $\vec p \vec p$ mode, it will be possible to measure with a
great precision  the double helicity PV asymmetry :
\EQ
\label{ALLPVdef}
A_{LL}^{PV} ={d\sigma_{(-)(-)}-d\sigma_{(+)(+)}\over 
d\sigma_{(-)(-)}+d\sigma_{(+)(+)}}
\eq
\noindent
where the signs  $\pm$ refer to the helicities of the colliding protons.
The cross section $d\sigma_{(\lambda_1)(\lambda_2)}$ means the one-jet
production cross section in a given helicity configuration, \linebreak
$p_1^{(\lambda_1)}p_2^{(\lambda_2)} \r jet + X$, estimated at  
$\sqrt{s} \ =\ 500$ GeV
for a given jet transverse energy \ET, integrated over a pseudorapidity interval 
$\Delta \eta \, =\, 1$ centered at $\eta\,=\,0$.  
\m
The \ZP will generate new amplitudes in quark-quark scattering which is
the dominant subprocess in the $E_T$ range we consider. At LO these amplitudes
will interfere with the one-gluon exchange amplitude and also with the amplitudes
due to SM gauge boson exchanges. This was described already in Ref. \cite{TVZprime}
and all the LO cross sections for the subprocesses can be found in Ref.
\cite{BouGuiSof}.   
\m
In fact, 95\% of the PV effect due to the new boson will come from
\ZP - gluon interference terms involving the scattering of $u$ quarks
in the $t$-channel.\\
In short notations :
\EQ
\label{ALLPVdom}
A_{LL}^{PV}. d\sigma \,\simeq\,  F
\int
\left((C_L^{u})^2 - (C_R^{u})^2 \right)
\biggl[u(x_1,\mu^2)\Delta u(x_2,\mu^2) + \Delta u(x_1,\mu^2)u(x_2,\mu^2) 
 \biggl]
\eq
\no
where $F$ is a positive factor  given by
\EQ
F \; =\; {32 \over 9} \alpha_s\,\alpha_Z
\hat s^2 Re \left( {1
\over \hat t D_{Z'}^{\hat u}} +
{1 \over \hat u D_{Z'}^{\hat t}}\right) 
\eq 
\no
where $\alpha_Z = \alpha/\sin^2\theta_W\cos^2\theta_W$ and
\EQ
D_{Z'}^{\hat t(\hat u)} \; =\; (\hat t(\hat u) - M_{Z'}^2) \, +\, 
i M_{Z'}\Gamma_{Z'}
\eq
In eq.(\ref{ALLPVdom}) $\Delta u(x,\mu^2) = u^+(x,\mu^2) - u^-(x,\mu^2)$ where
$u^{\pm}(x,\mu^2)$ are  the distributions of
the polarized $u$ quarks, either with helicity parallel (+) or
antiparallel (-) to the parent proton helicity. 
Summing $u^+$ and $u^-$ one recovers the unpolarized distribution
$u(x,\mu^2)$.
Concerning these
spin-dependent distributions (evaluated at $\mu = E_T$), we have used the
ones of  GRSV \cite{GRSV}. Note that the first part of the polarized RHIC 
Spin program
itself \cite{RSC} will greatly improve our knowledge of all the
distributions $\Delta q_i$'s and $\Delta \bar q_i$'s 
(see Ref. \cite{SofferVirey} for a recent analysis).
In the present analysis, the \ZP couplings to quarks remain in a "weak" range
(of the same order as the SM \Z couplings) : this is
at a variance with the phenomenological attempts of Refs.\cite{AltaChiap}. As a
consequence the \ZP width $\Gamma_{Z'}$  remains
in the range of 1 - 10 GeV/$c^2$ (neglecting possible light exotica), 
the precise value having no influence on our results. 
\m
We present in Fig.2 the result of our calculation  (taking all the terms,
dominant or not, into account) for \ALLPV versus $E_T$ in $\vec p \vec p$ collisions
at RHIC. For illustration we compare the SM asymmetry, which is due to small
QCD-electroweak interference terms \cite{AbudBaurGloverMartin}, 
to the non-standard one with a \ZP of mass
90 or 200 GeV$/c^2$  for models (A) and (B) and 300 GeV$/c^2$ for model (C).
The bump in the standard \ALLPV at $E_T \approx M_{Z,W}/2$ corresponds to the
vicinity of the corresponding "jacobian peak" in the production cross section. 
We observe the same behaviour at $E_T \approx 100$GeV for $M_{Z'} = 200$ GeV$/c^2$.
Practically, it will be difficult to explore the $E_T$ region below 45 GeV with the
RHIC detectors due to experimental cuts \cite{Epley}. 
However, above $E_T = 50$ GeV, on can see that a high precision measurement can be
performed. 
The error bars correspond to the statistical error with an integrated
luminosity ${\cal L}_1 = 800 pb^{-1}$ which can be achieved at RHIC in a few months
running. 
 
Clearly, if a deviation from the SM is seen in \ALLPV, the sign of this
deviation can allow to separate easily a model dominated by left-handed couplings
to $u$ quarks (like model (A)) from a model 
dominated by  right-handed couplings
like model (B) or (C). Of course the results are impressive because the mass
of the leptophobic \ZP is light. On the other hand, we have taken a value
of $\kappa$ around 1.5, when it is allowed by present data, or around 1 if not.
Both values are compatible with "weak" \ZP couplings to quarks (to be compared
with the models in \cite{AltaChiap,TVZprime}). 
\m
We have shown in Fig.1 the limits on the parameter space ($\kappa,M_{Z'}$) one
can obtain from \ALLPV with the integrated luminosities
${\cal L}_1 = 800 \, pb^{-1}$
and ${\cal L}_2 = 3200 \, pb^{-1}$ for models (A) and (B). One can see that the hole
centered on $M_{Z'} \approx 300 $GeV/$c^2$ is now fully covered, along with the low
mass or "degenerate" case $M_{Z'} \approx M_Z$.
Fig.3 is more general since it is model independent. It represents some exclusion
contours at 95\% C.L. in the plane ($\kappa.C_L$, $\kappa.C_R^{u}$) for two values
of the \ZP mass. The circles represent some estimates of the present experimental
constraints on this parameter space assuming some simple additional relations
 ($(C_R^{d})^2 = (C_R^{u})^2$)
between the couplings entering the dijet cross section. 
This assumption is not necessary to get the contours obtained from \ALLPV since
the $d$-quark couplings have a very small influence on this asymmetry.\\
The particular models (A), (B) and
(C) are represented by some particular points on the same plot for
$\kappa = 1$. Moving away from this $\kappa$ value results in a translation
along a straight line as shown. One can see that the measurement of \ALLPV
at RHIC will strongly reduce the allowed parameter space, in particular
in the region around $M_{Z'} = 300$ GeV$/c^2$ which is poorly constrained
up to now. Only when parity is restored, that is along the diagonals,
this kind of measurement cannot provide us some useful information.
\section{Conclusions}

Interest is growing on the potentialities offered by the polarized RHIC collider to
get a handle on new phenomena thanks to precision measurements of spin
asymmetries \cite{RIKENBNL}. We had first investigated in Ref. \cite{TVZprime} the
influence on \ALLPV in $\vec p \vec p$ collisions of
a purely phenomenological \ZP with strong couplings to quarks and a mass
around  1 TeV$/c^2$. Now it turns out that
various theoretical models are in favor of a lighter and leptophobic new neutral
gauge boson displaying chiral couplings to quarks. We have checked in this paper
that it is particularly relevant to search for some effects in the experimental
conditions of RHIC. In particular, the quite poor information available from
the  $p \bar p$ dijet experiments could be complemented for a wide part of the
parameter space. Moreover, in case of a positive signal
it is possible to get immediately an information on the chiral structure of the new
interaction. However it is still not possible to discriminate between 
a SUSY or a non-SUSY origin of the new \ZP.\\
Finally, it is interesting to note that a general prediction of all the leptophobic
models we have investigated (which assume a trilinear Yukawa term for
$u$-type quarks) is that $C_L^{u} \neq C_R^{u}$ since the Higgs doublet $H_U$ is
assumed to be charged under the group $U(1)'$. This implies that parity is violated
in the $u$-quark sector except for the very special axial case : $C_L^{u} = -
C_R^{u}$. On the other hand, parity is conserved for $d$-quark couplings in the
minimal two- Higgs doublets models \cite{BKMR12}, which is not a priori the case
for non-minimal models  (see e.g. Refs.\cite{CveticLang2,GG96}). 
Unfortunately, as we have said before, $d$-quark interactions are essentially masked
in  $\vec p \vec p$ collisions. It will be mandatory to perform polarized
neutron-neutron collisions to get a complementary information : at RHIC this could
be  realized with polarized $^3He$ beams as discussed recently
\cite{Courant}. Motivated by these arguments, and also by our earlier work
on the charged gauge boson sector \cite{TVWprimeBNL}, we are strongly in favor
of a complete polarization program at RHIC.

\m
\vspace*{2cm}
\no {\bf Acknowledgments}\\ 
J.M.V. acknowledges the warm hospitality at the RIKEN-BNL Research center where
part of this work has been performed. Thanks are due to 
G. Bunce, G. Eppley, B. Kamal, N. Saito, M. Tannenbaum and W. Vogelsang for fruitful
discussions. P.T. wishes to thank Yves Bigot for kind referencing.

\newpage

\newpage
{\bf Figure captions}
\bigbreak
\no
{\bf Fig. 1} Bounds on the parameter space ($\kappa,M_{Z'}$) in models A
and B. Contours are from the
dijet cross section analysis in $p \bar p$ collisions at CERN (UA2) (90\% C.L.)
and FNAL (CDF,D0) (95\% C.L.). The dotted (dashed) line corresponds to the predicted
limit  at (95\% C.L.) from 
\ALLPV in polarized $pp$ collisions at RHIC with a c.m. energy of 500 GeV and an
integrated luminosity ${\cal L}_1 = 800 pb^{-1}$ (${\cal L}_2 = 3200 pb^{-1}$).

\bigbreak
\no
{\bf Fig. 2}
\ALLPV for one-jet inclusive production, versus $E_T$, for polarized
$pp$ collisions at RHIC.  The plain curve is the SM expectation,
the dotted curves correspond to the "degenerate"
case $M_{Z'} = M_Z$ in model (A) (upper curve) and model (B) (lower curve).
Same for the dashed curves with $M_{Z'} = 200$ GeV$/c^2$. The dash-dotted curve 
corresponds to model (C) with $M_{Z'} = 300$ GeV$/c^2$.
The error bars correspond to the
statistical error with the luminosity ${\cal L}_1$. 

\bigbreak
\no
{\bf Fig. 3}
Contour plots at 95\%C.L. in the plane ($\kappa.C_L$, $\kappa.C_R^{u}$) from \ALLPV
measured with the integrated luminosity ${\cal L}_2$.
Plain (dashed) curves are for $M_{Z'} = 200 (300)$ GeV$/c^2$.
The circles correspond to exclusion limits from present collider data.
The black triangle, square and disk correspond to models (A), (B) and (C)
respectively for $\kappa = 1$.

\bigbreak \no

\bigbreak
\no
\newpage
\begin{figure}[ht]
\vspace{-1.5cm}
    \centerline{\psfig{figure=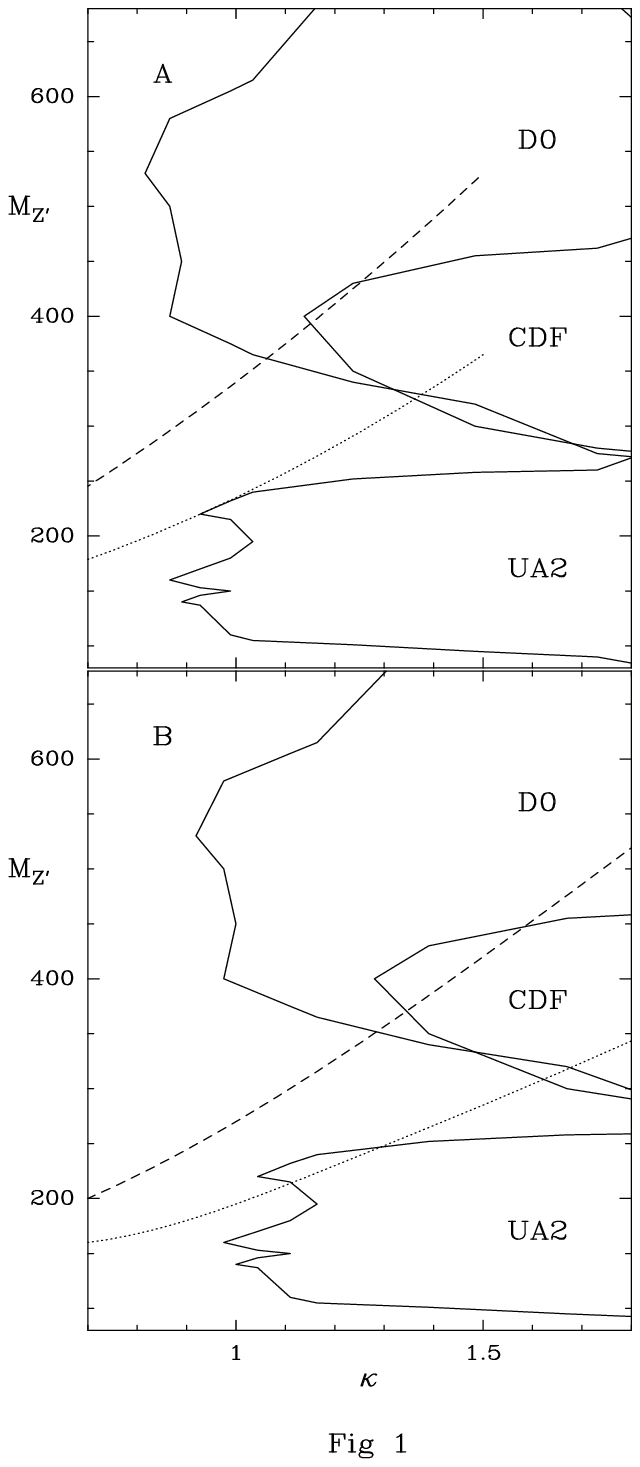,width=18cm}}
\end{figure}

\begin{figure}[ht]
\vspace{-2.5cm}
    \centerline{\psfig{figure=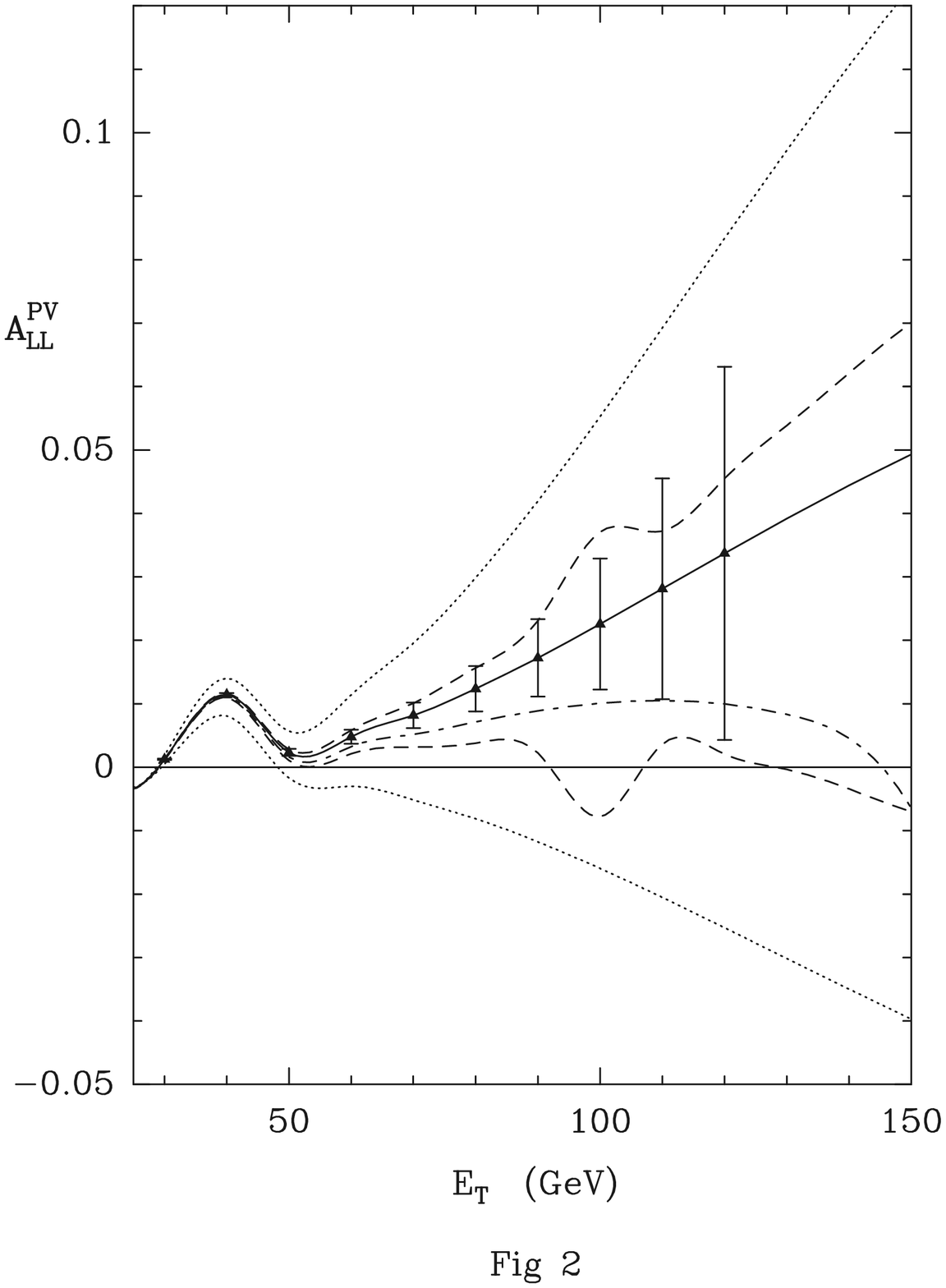,width=18cm}}
\end{figure}

\begin{figure}[ht]
\vspace{-2.5cm}
    \centerline{\psfig{figure=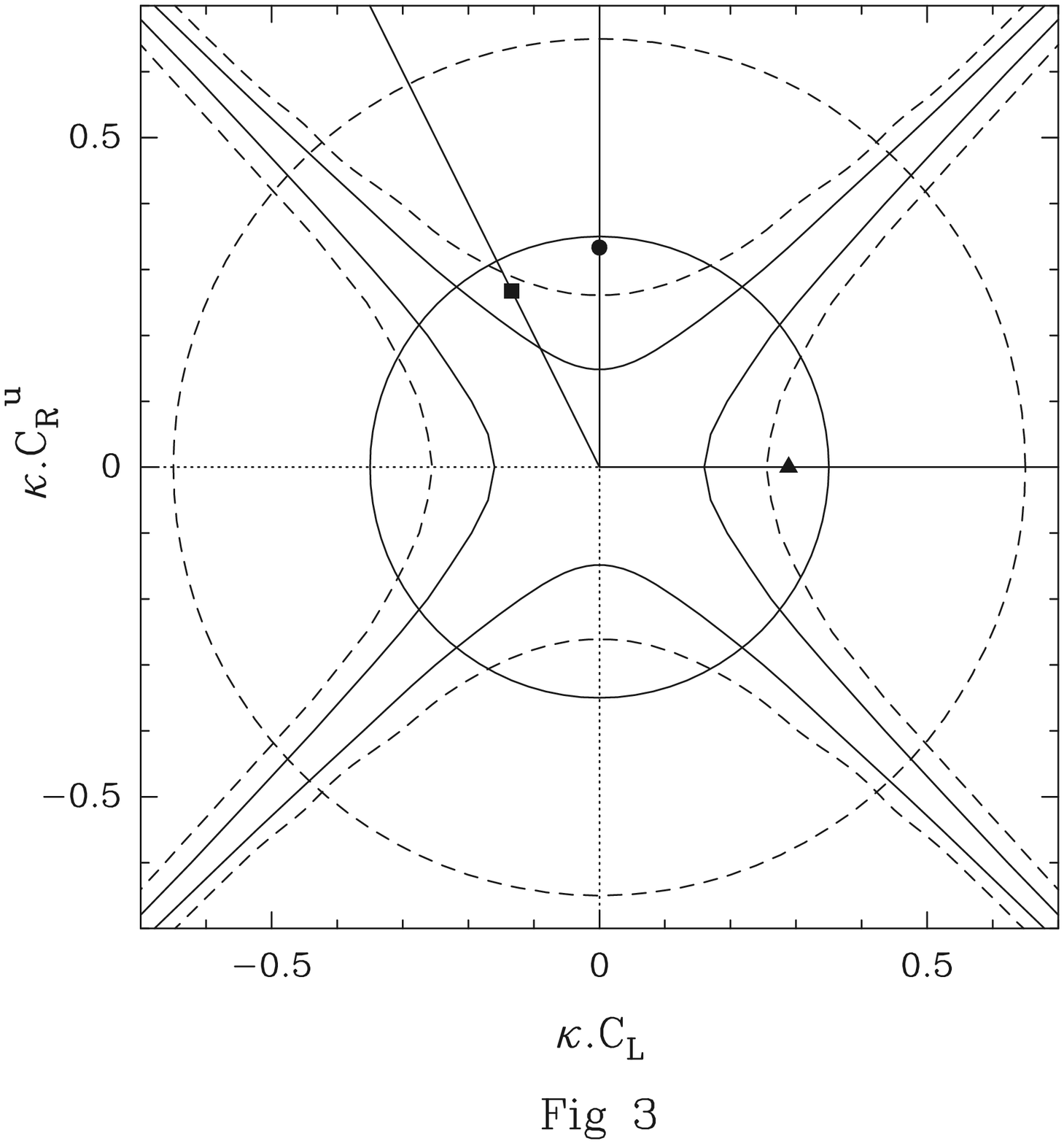,width=18cm}}
\end{figure}


\begin{thebibliography}{99}

\bibitem{LEPEWWG}
The LEP Collaborations Aleph, Delphi, L3, Opal, the LEP Electroweak Working
  Group and the SLD Heavy Flavor Group, CERN-PPE/97-154.

\bibitem{Mangano}
M.L. Mangano, proceedings of the Warsaw Conference ICHEP96, p. 1332.

\bibitem{CveticLang1}
M. Cveti{\v c} and P. Langacker, Phys. Rev. {\bf 54} (1996) 3570.

\bibitem{CveticLang2}
M. Cveti{\v c}, D.A. Demir, J.R. Espinosa, L. Everett and P. Langacker, Phys.
  Rev. {\bf 56} (1997) 2861.

\bibitem{Lykken2}
J. D. Lykken in Snowmass 1996, ed. D.G. Cassel, L. Trindle Gennari and R.H.
  Siemann, p. 891.

\bibitem{LangackerWang}
P. Langacker and J. Wang, hep-ph/9804438, see also P. Langacker hep-ph/9805486.

\bibitem{Rosner}
J.L. Rosner, Phys. Lett. {\bf B387} (1996) 113.

\bibitem{TannenbaumPenn}
M. Tannenbaum, in {\it Polarized Collider Workshop}, J. Collins, S.F.
  Heppelmann and R.W. Robinett eds, AIP Conf. Proceedings {\bf223}, AIP, New
  York, 1990, p. 201.

\bibitem{RSC}
G. Bunce et al. (RHIC Spin Collaboration), {\it Polarized protons at RHIC},
  Particle World, {\bf 3} (1992) 1 ; RSC, Letter of intent, April 1991 and RSC
  (STAR/PHENIX) letter of intent update, August 1992, BNL Reports, unpublished.

\bibitem{RIKENBNL}
Proceedings of the RIKEN-BNL research center workshop, april 1998, BNL Report
  65615.

\bibitem{AltaChiap}
G. Altarelli et al., Phys. Lett. {\bf B375} (1996) 292 ; P. Chiappetta et al.,
  Phys. Rev. {\bf D54} (1996) 789.

\bibitem{TVZprime}
P. Taxil and J.-M. Virey, Phys. Lett. {\bf B383}, 355 (1996).

\bibitem{LopezNanopoulos}
J.L. Lopez and D.V. Nanopoulos, Phys. Rev. {\bf D55} (1997) 397.

\bibitem{AEHN}
I. Antoniadis, J. Ellis, J.S. Hagelin and D.V. Nanopoulos, Phys. Lett. {\bf
  B194} (1987) 321.

\bibitem{BKMR12}
K.S. Babu, C. Kolda and J. March-Russell, Phys. Rev. {\bf D54} (1996) 4635 ;
  {\bf D57} (1998) 6788.

\bibitem{Hewett1}
J.L. Hewett and T.G. Rizzo, Phys. Reports, {\bf 183}, 193 (1989).

\bibitem{delA2}
F. del Aguila, G. Blair, M. Daniel and G.G. Ross, Nucl. Phys. {\bf B283} (1987)
  50.

\bibitem{FaraggiMasip}
A.E. Faraggi and M. Masip., Phys. Lett. {\bf B388} (1996) 524.

\bibitem{Agashe}
K. Agashe, M. Graesser, I. Hinchliffe and M. Suzuki, Phys. Lett. {\bf B385}
  (1996) 218.

\bibitem{GG96}
H. Georgi and S.L. Glashow, Phys. Lett. {\bf B387}, 341 (1996).

\bibitem{Hagiwara12}
Y. Umeda, G.C. Cho and K. Hagiwara, hep-ph/9805447, hep-ph/9805448.

\bibitem{UA2}
J. Alitti et al., Zeit. f. Phys. {\bf C49} (1991), 17 ; Nucl. Phys. {\bf B400}
  (1993), 3.

\bibitem{CDFWjets2}
F. Abe et al., Phys. Rev. {\bf D55} (1997) R5263.

\bibitem{D0jets}
B. Abbott et al., FERMILAB-Conf-97/356-E.

\bibitem{CaravagliosRoss}
F. Caravaglios and G.G.Ross, Phys. Lett. {\bf B346} (1995), 159.

\bibitem{BouGuiSof}
C. Bourrely, J. Ph. Guillet and J. Soffer, Nucl. Phys. {\bf B361} (1991) 72.

\bibitem{GRSV}
M. Gl\"uck, E. Reya, M. Stratman and W. Vogelsang, Phys. Rev. {\bf D53} (1996),
  4775.

\bibitem{SofferVirey}
J. Soffer and J.-M. Virey, Nucl. Phys. {\bf B509} (1998) 297.

\bibitem{AbudBaurGloverMartin}
M. Abud, R. Gatto and C.A. Savoy, Ann. Phys. (NY) {\bf 122} (1979) 219 ; U.
  Baur, E.W.N. Glover and A.D. Martin, Phys. Lett. {\bf B232} (1989) 519.

\bibitem{Epley}
G. Eppley in Ref.[10], p. 301.

\bibitem{Courant}
E. Courant in Ref.[10], p. 275.

\bibitem{TVWprimeBNL}
P. Taxil and J.-M. Virey, Phys. Lett. {\bf B404}, 302 (1997); J.-M. Virey in
  Ref.[10], p. 293.

\end{thebibliography}
\end{document}